\documentclass[aps,prl,twocolumn,showpacs]{revtex4}
\usepackage{amssymb}
\usepackage{amsmath}
\usepackage{graphicx}
\begin{document}
\title{Multistage Entanglement Swapping}
\author{Alexander M. Goebel$^1$}
\author{Claudia Wagenknecht$^1$}
\author{Qiang Zhang$^{2}$}
\author{Yu-Ao Chen$^{1,2}$}
\author{Kai Chen$^{2}$}\email{kaichen@ustc.edu.cn}
\author{J\"{o}rg Schmiedmayer$^{3}$}
\author{Jian-Wei Pan$^{1,2}$}

\affiliation{$^1$Physikalisches Institut,
Ruprecht-Karls-Universit\"{a}t Heidelberg, Philosophenweg 12,
69120 Heidelberg, Germany\\
$^2$Hefei National Laboratory for Physical Sciences at Microscale
and Department of Modern Physics, University of Science and
Technology of China, Hefei, Anhui 230026, China\\
$^3$Atominstitut der \"{o}sterreichischen Universit\"{a}ten, TU-Wien,
A-1020 Vienna, Austria}

\begin{abstract}
We report an experimental demonstration of entanglement swapping
over two quantum stages. By successful realizations of two cascaded
photonic entanglement swapping processes, entanglement is generated
and distributed between two photons, that originate from independent
sources and do not share any common past. In the experiment we use
three pairs of polarization entangled photons and conduct two
Bell-state measurements (BSMs) one between the first and second
pair, and one between the second and third pair. This results in
projecting the remaining two outgoing photons from pair 1 and 3 into
an entangled state, as characterized by an entanglement witness. The
experiment represents an important step towards a full quantum
repeater where multiple entanglement swapping is a key ingredient.
\end{abstract}

%\date{\today}
\date{May 30, 2008}
\maketitle

Entanglement swapping is arguably one of the most important
ingredients for quantum repeaters and quantum relays, which lays at
the heart of quantum communication
\cite{Zukowski1993,qrepeater,qrelay,QOreview}. For photonic quantum
communication, the distance is largely limited due to decoherence
from coupling to the environment and an increasing loss of photons
in a quantum channel. This leads to an exponential decay in the
fidelity of quantum information. This drawback can eventually be
overcome by subdividing larger distances into smaller sections over
which entanglement or quantum states can be distributed. The
sections are then bridged by entanglement swapping processes
\cite{qrepeater,qrelay}. The swapping procedure therefore
constitutes one of the key elements for a quantum relay
\cite{qrelay}, and a full quantum repeater \cite{qrepeater} if
combined with quantum purification
\cite{BDSW1996,purificationexperiments} and quantum memory
\cite{memory}. As a result, quantum communication becomes feasible
despite of realistic noise and imperfections. At the same time, the
overhead for the used resources and communication time only increase
polynomially with the distance \cite{qrepeater,qrelay,QOreview}.

Experimentally, photonic entanglement swapping has so far been
successfully achieved for the case of discrete variables
\cite{discreteswapping,Gisin2007}, and for continuous variable
\cite{CVswapping}, both via a single stage process. However,
only after successful multiple swapping, will we be able to
have a fully functional quantum repeater. There are
additional advantages utilizing a multiple swapping process. For a quantum
relay with many segments, it is equivalent to significantly lower
the dark-count rate, which is a substantial factor limiting the
transmission distance of successful quantum communication
\cite{qrelay}. For quantum information carriers possessing mass,
multiple swapping processes can speed up the distribution of
entanglement by a factor that is proportional to the number of
segments used \cite{multiparticleswapping}. Moreover, multistage
entanglement swapping can improve the protection of quantum states
against noise from amplitude errors
\cite{multiparticleswapping}.

We report in this letter an experimental demonstration of a multiple
entanglement swapping over two stages. This is achieved by utilizing
three synchronous spatially independent pairs of polarization
entangled photons, and performing BSMs among the three segments
between the two communication parties. Two successful BSMs yield a
final maximally entanglement pair distributed between the two
parties. To quantitatively evaluate the performance, we have
observed the quality of the output state by the characterization of
an entanglement witness, which confirms genuine entanglement
generation. Our experiment implements an entanglement distribution
over two distant stations which are initially independent
of each other and have never physically interacted in the past. This
proof-of-principle demonstration constitutes an important step
towards robust long-distance quantum relays, quantum repeaters and
related quantum protocols based on multiple entanglement swapping.

\begin{figure}[ptb]
\begin{center}
\includegraphics[width=3in ]{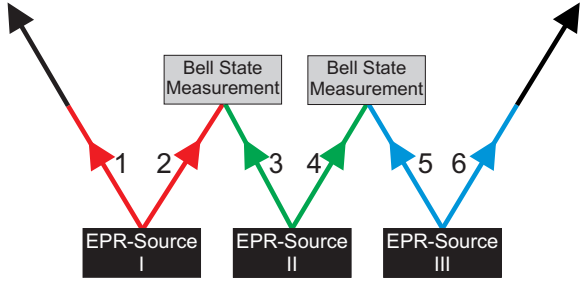}
\end{center}
\caption{Principle of multistage entanglement swapping: three EPR
sources produce pairs of entangled photons 1-2, 3-4 and 5-6. Photon
2 from the inial state and photon 3 from the first ancillary pair
are subjected to a joint BSM, and so are photon 4 from the first
ancillary and photon 5 from the second acillary pair. The two BSMs
project outgoing photons 1 and 6 onto an entangled state. Thus the
entanglement of the initial pair is swapped to an entanglement
between photons 1 and 6.} \label{sketchofprinciple}
\end{figure}

The principle for multistage entanglement swapping is sketched in
Fig. \ref{sketchofprinciple}. Consider three independent stations,
each simultaneously emitting a pair of Einstein-Podolsky- Rosen
(EPR) maximally entangled photons. In our experiments, we generate
these states through the process of spontaneous parametric
down-conversion \cite{SPDCII}. By post-selecting events with only
one photon in each output arm, we obtain polarization entangled
photons in the state
\begin{equation}
|\Psi\rangle_{123456} =
|\Psi^{-}\rangle_{12}\times|\Psi^{-}\rangle_{34}\times|\Psi^{-}\rangle_{56},
\label{groundstate}
\end{equation}
where $|\Psi^{-}\rangle_{ij}$ is one of the four maximally entangled
Bell states, which form a complete orthonormal basis for the joint
state of two entangled photons
\begin{eqnarray*}
|\Psi^{\pm}\rangle_{ij} & = &
\tfrac{1}{\sqrt{2}}(|H\rangle_{i}|V\rangle_{j}\pm|V\rangle_{i}|H\rangle_{j})
\nonumber\\
|\Phi^{\pm}\rangle_{ij} & = &
\tfrac{1}{\sqrt{2}}(|H\rangle_{i}|H\rangle_{j}\pm|V\rangle_{i}|V\rangle_{j}).
\label{Bellstates}
\end{eqnarray*}
Here $|H\rangle$ ($|V\rangle$) denotes the state of a horizontally
(vertically) polarized photon. Note that photon pairs 1-2, 3-4 and
5-6 are entangled in an antisymmetric polarization state. The states
of the three pairs are factorizable from each other, namely, there
is no entanglement among photons from different pairs.

As a first step we perform a joint BSM on photons 2 and 3, that is,
photons 2 and 3 are projected onto one of the four Bell states.
This measurement also projects photons 1 and 4 onto a Bell
state, in a form depending on the result of the BSM of photons 2 and 3. Close
inspection shows that for the initial state given in
Eq.~(\ref{groundstate}), the emerging state of photons 1 and 4 is
identical to the one that photons 2 and 3 collapse into. This is a
consequence of the fact that the state of Eq.~(\ref{groundstate})
can be rewritten as
\begin{eqnarray}
|\Psi\rangle_{123456} & = &\tfrac{1}{2}\,
[|\Psi^{+}\rangle_{14}|\Psi^{+}\rangle_{23}
-|\Psi^{-}\rangle_{14}|\Psi^{-}\rangle_{23}\nonumber\\
&&-|\Phi^{+}\rangle_{14}|\Phi^{+}\rangle_{23}
+|\Phi^{-}\rangle_{14}|\Phi^{-}\rangle_{23}]\nonumber\\
&&\times|\Psi^{-}\rangle_{56}
\label{totalstate}
\end{eqnarray}
In all cases photons 1 and 4 emerge entangled despite the fact that
they never interacted with one another in the past.
The joint measurement of photons 2 and 3 tells about the type of
entanglement between photons 1 and 4.

Without loss of generality, we assume in the first step that
photons 2 and 3 have collapsed into the state
$|\Phi^{+}\rangle_{23}$ as a result of the first BSM. The remaining
four-photon state is then of the form
\begin{eqnarray}
|\Psi\rangle_{1456} & = &\tfrac{1}{2}\,
[|\Psi^{+}\rangle_{16}|\Phi^{-}\rangle_{45}
+|\Psi^{-}\rangle_{16}|\Phi^{+}\rangle_{45}\nonumber\\
&&-|\Phi^{+}\rangle_{16}|\Psi^{-}\rangle_{45}
-|\Phi^{-}\rangle_{16}|\Psi^{+}\rangle_{45}]
\label{state1456}
\end{eqnarray}

In a similar manner we perform a second BSM on
photons 4 and 5. Again a detection of the state
$|\Phi^{+}\rangle_{45}$ results in projecting the remaining photons
1 and 6 onto the Bell state
\begin{equation}
|\Psi^-\rangle_{16}=\tfrac{1}{\sqrt{2}}(|H\rangle_{1}|V\rangle_{6}-
|V\rangle_{1}|H\rangle_{6})
\label{finalstate16}
\end{equation}

%%%%%%%%%%%%%%%%%%%%%%%%%%%%%%%%%%%%%%%%%%%%%%%%%%%%%%%%%%%
%description of experimental setup
%%%%%%%%%%%%%%%%%%%%%%%%%%%%%%%%%%%%%%%%%%%%%%%%%%%%%%%%%%%

%\begin{figure}[htb]
\begin{figure}[t]
\begin{center}
\includegraphics[width=3in]{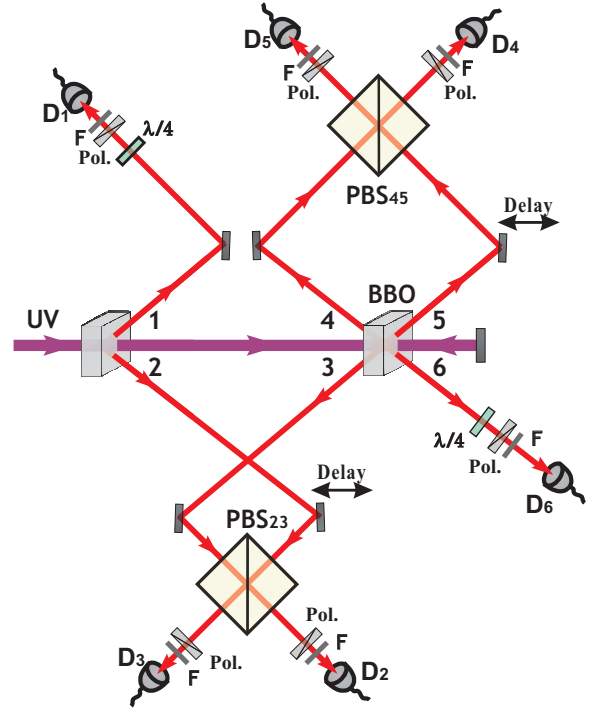}
\end{center}
\caption{The focused ultraviolet laser beam passes the first BBO
generating photon pair 1-2. Refocussed, it passes the second BBO
generating the ancillary pair 5-6 and again retroreflected through
the second BBO generating pair 3-4. In order to achieve
indistinguishability at the interference PBS23 and PBS45 the spatial
and temporal overlap are maximized by adjusting the delays and
observing `Shih-Alley-Hong-Ou-Mandel'-type interference fringes
\cite{Shih-Alley-Hong-Ou-Mandel} behind the PBS23 (PBS45) in the
$\pm$ basis \cite{Pan2001}.
With the help of polarizers and half/quarter wave plates, we are
able to analyze the polarization of photons in arms 1 and 6.
All photons are spectrally filtered by
narrow band filters with $\Delta \lambda_{\text{FWHM}}\approx 2.8
\text{nm} $ and are monitored by silicon avalanche single-photon
detectors \cite{Zukowski1995}. Coincidences are counted by a laser
clocked field-programmable gate array based coincidence unit.}
\label{setupswap}
\end{figure}
A schematic diagram of our setup for multistage entanglement
swapping is illustrated in Fig.~\ref{setupswap}. We use a pulsed
high-intensity ultraviolet (UV) laser with a central wavelength of
390nm, a pulse duration of around 180 fs and a repetition rate of 76
MHz. The beam successively passes through two $\beta$-Barium-Borate
(BBO) crystals, and is reflected to pass again through the second
BBO to generate three polarization entangled photon pairs via
type-II parametric down conversion \cite{SPDCII}.

Due to the high average power of 1W UV-light and improvements in
collection efficiency and stability of the photon sources
\cite{Zhang}, we are able to observe up to $10^5$ photon pairs per
second from each source. With this brightness of the entangled
photon sources we could obtain around 4.5 six-photon events per
minute in our setup.

For the joint BSM of photons 2 and 3 (photons 4 and 5), we choose to
analyze the case of detecting the projection onto a
$|\Phi^{+}\rangle$ state. Using a polarizing beam splitter (PBS)
allows the projection of photons 2 and 3 (4 and 5) onto the state
$|\Phi^{+}\rangle$ upon detecting a $|+\rangle|+\rangle$ or
$|-\rangle|-\rangle$ coincidence at detectors D2 and D3 (D4 and D5)
(with $|\pm\rangle=(|H\rangle\pm|V\rangle)/\sqrt{2}$). In our
experiment only the $|+\rangle|+\rangle$ coincidences were
registered, which reduces the overall success probability by a
factor of 1/64. This could be improved by installing a half wave
plate (HWP) at $22.5^{\circ}$, which corresponds to a polarization
rotation of $45^{\circ}$, and a PBS after each output arm of PBS23
(PBS45). This configuration would also allow to detect the state
$|\Phi^{-}\rangle$, which results in a $|+\rangle|-\rangle$ or
$|-\rangle|+\rangle$ coincidence \cite{Pan1998}. Thus, a factor of
1/4 for the overall success probability could be achieved in an
ideal case.

%%%%%%%%%%%%%%%%%%%%%%%%%%
%  presentation of data  %
%%%%%%%%%%%%%%%%%%%%%%%%%%

As shown in equations
Eq.~(\ref{totalstate},\ref{state1456},\ref{finalstate16}) the
projection measurements onto $|\Phi^{+}_{23}\rangle$ and
$|\Phi^{+}_{45}\rangle$ leave photons 1 and 6 in the maximally
entangled state $|\Psi^{-}_{16}\rangle$. In contrast to quantum
state tomography, the measurement of witness operators does not
provide a complete reconstruction of the original quantum state, it
however allows to check with a minimal number of local measurements
for a entanglement character of a quantum state. To verify that the
two photons are really in an entangled state, and thus the swapping
operation is successful, the expectation value of the corresponding
witness operator \cite{Guehne2002,Barbieri2003} is expected to take
a negative value. In our case, the applied witness operator $W$ is
the most efficient one since it involves only the minimal number of
local measurements \cite{Guehne2002}. It can be measured locally by
choosing correlated measurement settings, that involve only the
simultaneous detection of linear, diagonal and circular
polarizations for both photons. We have performed local measurements
on the outgoing state of photons 1 and 6 in the three complementary
bases; linear (H/V), diagonal (+/-) and circular (R/L) (with
$|L\rangle=(|H\rangle+i|V\rangle)/\sqrt{2}$ and
$|R\rangle=(|H\rangle-i|V\rangle)/\sqrt{2}$).

The entanglement witness is given by
\begin{eqnarray}
W& = & \tfrac{1}{2}\ (|HH\rangle\langle HH|
+|VV\rangle\langle VV| + |++\rangle\langle ++| \nonumber\\
&& + |--\rangle\langle --| - |RL\rangle\langle RL|
-|LR\rangle\langle LR|). \label{witnessform}
\end{eqnarray}
\begin{figure}[ptb]
\begin{center}
\includegraphics[width=3in]{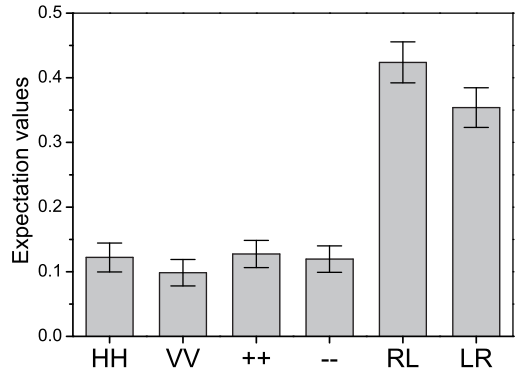}
\end{center}
\caption{Experimental expectation values for every correlation
function of the entanglement witness for the swapped state. The
results are derived by twofold coincidence measurements along
three complementary common bases: a) $|H\rangle|V\rangle$; b)
$|+\rangle|-\rangle$; and c) $|R\rangle|L\rangle$, conditioned on a
fourfold coincidence event in $|++++\rangle$ for detectors
D2-D3-D4-D5 which ensures two successful Bell state measurements.} \label{witness}
\end{figure}
In the experiment, we perform measurements for each correlation
function of the witness. The expectation values are
shown in Fig.~\ref{witness}. Experimental integration time for each
local measurement took about 60 hours and we recorded about 180
events of desired two-qubit coincidences. Every expectation value
for a correlation function is obtained by making a von Neumann
measurement along a specific basis and computing the probability over
all the possible events. For example, for a HH correlation
$\text{Tr} (\rho |HH\rangle\langle HH|)$, we perform measurements
along the H/V basis. Then its value is given by the number of
coincidence counts of HH over the sum of all coincidence counts of
HH, HV, VH and VV. We proceed likewise for the other correlation
settings. The witness can then directly be evaluated to
$\text{Tr}(\rho W)=-0.16 \pm 0.03$. The negativity of the measured
witness implies clearly that entanglement has indeed
been swapped. The imperfection of our data is due to the non-ideal
quality of entangled states generated from the high power UV beam,
as well as the partial distinguishability of independent photons at
PBS23 and PBS45, which leads to non-perfect interferences and a
degrading of entanglement output quality \cite{degrading}. Moreover,
double pair emission by a single source causes noise of an order of
10 spurious six-fold coincidences in 60 hours and was not subtracted
in calculating the expectation value of the witness operator.

\begin{figure}[ptb]
\begin{center}
\includegraphics[width=3in]{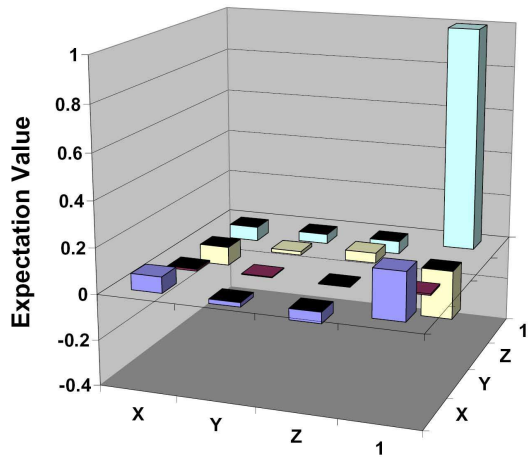}
\end{center}
\caption{Complete quantum state tomography on photon 1 and 6 before
entanglement swapping. Label X corresponds to measurement setting
$\sigma_x$, while Y and Z are for $\sigma_y$ and $\sigma_z$,
respectively. The result shows that the photons didn't reveal any
entanglement whatsoever before the swapping operation.}
\label{tomography}
\end{figure}

To ensure that there is no entanglement between photons 1 and 6
before either of the entanglement swapping process, we have
performed a complete quantum state tomography. The experimental
expectation values for various bases are illustrated in
Fig.~\ref{tomography}. Concurrence \cite{concurrence} is a monotonic
function of entanglement, ranging from 0 for a separable state to 1
for a maximally entangled state. In terms of concurrence, we can
thus quantify the degree of entanglement through a reconstructed
density matrix $\rho_{init}$ for the initial combined state from the
data shown in Fig.~\ref{tomography}. The concurrence $C_{init}$
derived from $\rho_{init}$ is $C_{init}=\max(0,-0.39 \pm 0.01)=0$.
As expected the concurrence is indeed 0, therefore photons 1 and 6
did not reveal any entanglement whatsoever before the swapping.
Ideally, for a completely mixed state the expectation values for all
local measurements should be 0, except for the unity operator, which
should be 1. The contributions of the measurement settings other
than the unity operator are mainly due to noise caused by scattered
light of the UV beam at the BBO crystal.
For convenience of comparison, we also performed the same witness
measurement of Eq.~(\ref{witnessform}) to give $\langle W \rangle=0.28\pm 0.01$,
which is safely above the bound $\langle W \rangle < 0$ needed to reveal entanglement.
However, after the two-stage entanglement swapping, entanglement arises as unambiguously
confirmed by negativity of expectation value for
the witness $\langle W \rangle =-0.16 \pm 0.03$  as discussed above.

In conclusion, we have for the first time provided a
proof-of-principle demonstration of a two-stage entanglement
swapping using photonic qubits. The feasibility and effectiveness of
this process has been verified by a successful distribution of
genuine entanglement after two simultaneously independent swapping
process. This result yields the possibility of immediate near-future
applications of various practical quantum information processing
tasks. If combined with narrow-band entanglement sources, the
implementation of quantum relays (without quantum memory) and
quantum repeaters (with quantum memory) would become within current
reach \cite{qrepeater,memory,Gisin2007}, as well as quantum state
transfer and quantum cryptography networks in a more efficient way
and over much larger distances of around hundreds of kilometers
\cite{qrelay}. Our demonstration also allows for the possibility of
utilizing multi-party, multiple stages entanglement swapping to
achieve global quantum communication networks though with
significant challenges ahead \cite{multiparticleswapping}.

This work was supported by the Marie Curie Excellence Grant from the
EU, the Alexander von Humboldt Foundation, the National Fundamental
Research Program of China under Grant No.2006CB921900, the CAS, and
the NNSFC. K.C acknowledges support of the Bairen program of CAS.
C.W. was additionally supported by the Schlieben-Lange Program of
the ESF.

\end{document}